\documentclass{ifacconf}

\usepackage{graphicx}
\usepackage{natbib}
\usepackage{siunitx}
\usepackage{amsmath,amssymb,amsfonts}
\newenvironment{proof}[1][Proof]{\par\noindent\textit{#1. }\ignorespaces}{\hfill$\square$\par}
\usepackage[ruled,lined,algo2e]{algorithm2e}
\usepackage{booktabs}
\usepackage[utf8]{inputenc}
\usepackage{newunicodechar}
\usepackage{pifont}
\newunicodechar{✗}{\ding{55}}
\usepackage{multirow}
\newunicodechar{✓}{\checkmark}
\newunicodechar{α}{\ensuremath{\alpha}}
\usepackage{enumitem}
\newunicodechar{δ}{\ensuremath{\delta}}
\newtheorem{assumption}{Assumption}
\newtheorem{problem}{Problem}
\newtheorem{proposition}{Proposition}
\newtheorem{definition}{Definition}
\newtheorem{remark}{Remark}
\newtheorem{theorem}{Theorem}

\begin{document}
\begin{frontmatter}

\title{AURORA: \textit{A}utonomous \textit{U}pdating of \textit{RO}M and Controller via \textit{R}ecursive \textit{A}daptation}

\author[First]{Jiachen Li}
\author[First]{Shihao Li}
\author[First]{Dongmei Chen}

\address[First]{University of Texas at Austin,
   Austin, TX 78712 USA (e-mail: jiachenli@utexas.edu, shihaoli@utexas.edu, dmchen@me.utexas.edu).}

\begin{abstract}
Real-time model-based control of high-dimensional nonlinear systems presents severe computational challenges. Conventional reduced-order model (ROM) control relies heavily on expert tuning or parameter adaptation and seldom offers mechanisms for online supervised reconstruction. We introduce AURORA (\textbf{A}utonomous \textbf{U}pdating of \textbf{RO}M and Controller via \textbf{R}ecursive \textbf{A}daptation), a supervisory framework that automates ROM-based controller design and augments it with diagnostic-triggered structural adaptation. Five specialized agents collaborate through iterative generate–judge–revise cycles, while an Evaluation Agent classifies performance degradation into three operationally distinct categories—subspace inadequacy, parametric drift, and control inadequacy—and routes corrective action to the responsible agent. For linear ROMs, we analytically prove that this classification is correct under mild assumptions and that the supervisory switching cycle preserves exponential stability subject to a dwell-time condition. For nonlinear systems, the absence of a universal Lyapunov construction for autonomously discovered ROM structures precludes analogous analytical guaranties; we therefore validate the same classification empirically. Experiments on eight benchmark systems with state dimensions up to $N = 5177$ compare AURORA against expert-tuned baselines, gain-scheduled control, and online RLS-adaptive alternatives. Controlled fault-injection experiments confirm 91\% diagnostic routing accuracy. AURORA achieves $6$--$12\%$ tracking improvement over expert baselines and $4$--$5\%$ over classical adaptive alternatives.
\end{abstract}


\end{frontmatter}

\section{Introduction}
\label{sec:introduction}

Controlling high-dimensional dynamical systems remains a fundamental computational challenge. Systems arising from the spatial discretization of PDEs, large-scale mechanical assemblies, or networked dynamics routinely involve state dimensions ranging from thousands to millions. Model-based control methods—including MPC \citep{rawlings2017model}, feedback linearization \citep{isidori1995nonlinear}, and optimal control \citep{kirk2004optimal}—require repeated evaluations of the governing dynamics and become computationally intensive at these scales.

Reduced-order modeling relieves this burden by exploiting the low-dimensional structure inherent in many physical systems while still capturing the major transient behavior of the systems. \citep{benner2015survey,brunton2019data}. Established projection-based techniques---POD \citep{sirovich1987turbulence}, balanced truncation \citep{benner2015survey}, and Galerkin projection \citep{holmes2012turbulence}---have matured considerably. Yet ROM-based control pipelines confront persistent obstacles: limited validity under parameter variation, parametric uncertainty, and projection error that accumulates during closed-loop operation. More fundamentally, existing pipelines demand expert judgment at every stage—method selection, order determination, and controller tuning. The existing pipelines lack any principled mechanism for supervisory redesign when operating conditions shift.

When parameters drift or the system trajectory departs from the subspace spanned by the ROM basis, the standard engineering response relies on manual re-identification and controller retuning—a process that demands specialized expertise and typically consumes several hours per system. Classical adaptive control \citep{narendra1989stable,astrom2013adaptive} can update parameters within a fixed model class but cannot restructure the ROM itself. What remains missing is an automated supervisory layer that diagnoses the source of degradation and dispatches the appropriate corrective action. This paper aims to fill this gap through the use of large language models (LLMs).

Recent advances in LLMs have opened new avenues for autonomous engineering design \citep{li2025generative, zhang2025large}, including controller tuning \citep{guo2024controlagent} and code generation with iterative refinement \citep{shinn2023reflexion,madaan2024self}. These capabilities suggest a practical route toward the supervisory layer just described: an LLM-based system that selects ROM methods, generates and validates controller code, and dispatches corrective redesign—all from a natural-language problem description. This particular approach—LLM--based supervisory automation for ROM--controller co-design—has yet to be explored. 

AURORA departs from ControlAgent \citep{guo2024controlagent}, which automates classical controller tuning for fixed models, and from ROM-MPC approaches \citep{carlberg2015galerkin,ahmed2024adversarially}, which presuppose manual ROM construction. By contrast, AURORA automates the entire pipeline—from a natural-language problem description to a deployed controller with supervisory adaptation.

\textbf{Contributions:} We propose AURORA, a supervisory framework for autonomous ROM-based controller design and adaptation. The contributions are organized according to the regime in which they apply:
\begin{enumerate}[leftmargin=*,itemsep=1pt]
    \item \textbf{Diagnostic classification with formal guaranties (linear ROM):} We formalize a classification of ROM-based controller degradation into three operationally distinct categories: subspace inadequacy, parametric drift, and control inadequacy. We prove identifiability, i.e., that the three categories are mutually exclusive and can be uniquely determined from monitored signals, under mild assumptions (Proposition~\ref{prop:identifiability}). We further establish that the supervisory switching cycle preserves exponential stability under a dwell-time condition—a minimum time between consecutive controller switches that prevents instability from overly rapid switching \citep{morse1996supervisory}—(Theorem~\ref{thm:switching}).
    \item \textbf{Empirical validation for nonlinear systems:} For nonlinear ROMs, we apply the same classification as a heuristic and validate it through controlled fault-injection experiments, achieving 91\% routing accuracy across five systems (Section~\ref{sec:diagnostic_validation}).
    \item \textbf{Systematic benchmark against classical alternatives:} We compare AURORA against expert-tuned baselines, gain-scheduled control \citep{rugh2000research}, online RLS-adaptive control \citep{astrom2013adaptive}, and a scripted non-LLM pipeline across eight benchmark systems spanning linear, nonlinear, and PDE dynamics (Section~\ref{sec:experiments}).
\end{enumerate}

The paper proceeds as follows. Section~\ref{sec:related} surveys related work. Section~\ref{sec:problem} formalizes the problem setting. Section~\ref{sec:aurora} presents the AURORA framework, covering agent design, diagnostic classification, and supervisory stability analysis. Section~\ref{sec:experiments} reports experimental results, diagnostic validation, and ablation studies. Section~\ref{sec:conclusion} concludes.

\section{Related Work}
\label{sec:related}

\textbf{ROM-Based Control.}
Reduced-order modeling is indispensable for real-time control whenever full-order dynamics are too costly to evaluate online. Physics-based ROMs have been developed over the past years. \cite{carlberg2015galerkin} develops structure-preserving Galerkin projection for nonlinear model reduction; \cite{ahmed2024adversarially} proposes adversarially robust ROM-MPC formulations. Data-driven alternatives such as SINDy \citep{brunton2016discovering} and DMD \citep{brunton2019data} infer low-dimensional dynamics directly from measurements. A persistent limitation across these methods is that ROM construction is treated as an offline procedure, with no supervisory mechanism for autonomous redesign once the system departs from nominal conditions.

\textbf{Supervisory and Adaptive Control.}
Supervisory control architectures \citep{morse1996supervisory,hespanha2003overcoming} switch among pre-designed controllers based on online performance monitoring. Classical adaptive control \citep{narendra1989stable,astrom2013adaptive} updates parameters within a fixed model class via Lyapunov-based or recursive estimation laws. More recent work—adaptive MPC with recursive constraint tightening \citep{lorenzen2017adaptive}, safe Bayesian learning for nonlinear MPC \citep{buerger2024safe}, and gain-scheduled MPC \citep{rawlings2017model}—addresses parameter uncertainty within model-based frameworks. AURORA differs in a key respect: it can restructure the ROM itself—enriching the basis or changing the reduction method—rather than merely adapting parameters within a pre-specified model class. The supervisory layer is consequently more general than classical adaptive control, though it operates on a slower timescale.

\textbf{LLM-Based Automation for Engineering.}
LLM-based agents have demonstrated broad capabilities in planning, reasoning, and tool use \citep{wang2024survey}, and have been extended to multi-agent architectures \citep{wu2023autogen,hong2023metagpt} and iterative self-refinement \citep{shinn2023reflexion,madaan2024self}. In control engineering specifically, \cite{guo2024controlagent} automate controller tuning through LLMs iteration, and \cite{liang2024lmpc} integrate LLMs with MPC for robotic manipulation. These efforts remain confined to classical controllers or full-order models and do not address ROM–controller co-design or supervisory redesign.

\section{Problem Formulation and Background}
\label{sec:problem}

\subsection{Problem Setting}

Consider a parameterized high-dimensional dynamical system
\begin{equation}
\dot{\mathbf{x}}(t) = \mathbf{f}(\mathbf{x}(t), \mathbf{u}(t); \boldsymbol{\theta}), \quad \mathbf{y}(t) = \mathbf{h}(\mathbf{x}(t))
\label{eq:fom}
\end{equation}
where $\mathbf{x} \in \mathbb{R}^N$ is the state, with $N$ potentially very large; $\mathbf{u} \in \mathcal{U} \subseteq \mathbb{R}^m$ is a constrained input; $\mathbf{y} \in \mathbb{R}^p$ is the measured output; and $\boldsymbol{\theta} \in \Theta \subseteq \mathbb{R}^{n_\theta}$ collects uncertain or time-varying parameters. These parameters may drift as $\boldsymbol{\theta}(t) = \boldsymbol{\theta}_0 + \delta\boldsymbol{\theta}(t)$ with $\|\delta\boldsymbol{\theta}\| \leq \varepsilon_\theta$.

In the linear special case, \eqref{eq:fom} reduces to $\dot{\mathbf{x}} = \mathbf{A}(\boldsymbol{\theta})\mathbf{x} + \mathbf{B}(\boldsymbol{\theta})\mathbf{u}$, $\mathbf{y} = \mathbf{C}\mathbf{x}$. The remainder of this section presents reduction and control techniques separately for the linear and nonlinear settings, reflecting the distinct theoretical tools available in each case.

\begin{problem}[Autonomous ROM-Based Control with Supervisory Adaptation]
\label{prob:main}
Given system~\eqref{eq:fom} described in natural language, together with constraints $\mathbf{u} \in \mathcal{U}$, $\mathbf{x} \in \mathcal{X}$, and a reference trajectory $\mathbf{y}_{\mathrm{ref}}(t)$:
\begin{enumerate}[leftmargin=*,itemsep=1pt]
    \item Autonomously select and construct a ROM $\dot{\mathbf{r}} = \mathbf{f}_r(\mathbf{r}, \mathbf{u})$ with $\mathbf{r} \in \mathbb{R}^r$, $r \ll N$;
    \item Design a controller $\pi_r$ on the ROM that achieves tracking $\|\mathbf{y} - \mathbf{y}_{\mathrm{ref}}\| \leq \varepsilon_{\mathrm{track}}$ while respecting all constraints;
    \item Provide a supervisory mechanism that monitors closed-loop performance, diagnoses the source of any degradation, and dispatches corrective redesign---without human intervention.
\end{enumerate}
\end{problem}

\subsection{Linear Reduction and Control}
\label{sec:linear_rom_ctrl}

\textbf{Projection-Based Reduction.}
Given a reduced basis $\mathbf{\Phi} \in \mathbb{R}^{N \times r}$ with $\mathbf{\Phi}^T\mathbf{\Phi} = \mathbf{I}_r$, the full state is approximated as $\mathbf{x}(t) \approx \mathbf{\Phi}\mathbf{r}(t)$. Galerkin projection yields the reduced dynamics $\dot{\mathbf{r}} = \mathbf{A}_r\mathbf{r} + \mathbf{B}_r\mathbf{u}$, $\mathbf{y} \approx \mathbf{C}_r\mathbf{r}$, where $\mathbf{A}_r = \mathbf{\Phi}^T\mathbf{A}\mathbf{\Phi}$, $\mathbf{B}_r = \mathbf{\Phi}^T\mathbf{B}$, and $\mathbf{C}_r = \mathbf{C}\mathbf{\Phi}$. POD \citep{sirovich1987turbulence} constructs $\mathbf{\Phi}$ from snapshot data via SVD, retaining modes that capture a prescribed fraction of the system energy ($\geq 99.5\%$). Balanced truncation \citep{benner2015survey} instead selects states that are simultaneously controllable and observable, ranked by Hankel singular values, and preserves stability by construction.

\textbf{Linear Control.}
For linear ROMs, LQR minimizes $J = \int_0^\infty (\mathbf{r}^T\mathbf{Q}_r\mathbf{r} + \mathbf{u}^T\mathbf{R}\mathbf{u})\,dt$ via the algebraic Riccati equation, where $\mathbf{Q}_r = \mathbf{\Phi}^T\mathbf{Q}_x\mathbf{\Phi}$ maps full-order cost objectives into reduced coordinates.

\subsection{Nonlinear Reduction and Control}
\label{sec:nonlinear_rom_ctrl}

\textbf{Nonlinear Reduction.}
For nonlinear systems, Galerkin projection gives $\dot{\mathbf{r}} = \mathbf{\Phi}^T\mathbf{f}(\mathbf{\Phi}\mathbf{r}, \mathbf{u})$, but evaluating $\mathbf{f}(\mathbf{\Phi}\mathbf{r})$ still incurs the cost of the full-order model. DEIM \citep{chaturantabut2010nonlinear} addresses this bottleneck by approximating the nonlinear terms at a small set of interpolation points, reducing computational cost to $O(r)$. As an alternative, SINDy \citep{brunton2016discovering} identifies parsimonious reduced equations $\dot{\mathbf{r}} = \boldsymbol{\Theta}(\mathbf{r}, \mathbf{u})\boldsymbol{\Xi}$ through sparse regression over a candidate function library.

\textbf{Nonlinear MPC on ROMs.}
When the system is nonlinear or subject to hard constraints, MPC solves:
\begin{equation}
\min_{\{\mathbf{u}_i\}_{i=0}^{N_p-1}} \sum_{i=0}^{N_p-1} \ell(\mathbf{r}_{k+i}, \mathbf{u}_{k+i}) + V_f(\mathbf{r}_{k+N_p})
\label{eq:mpc}
\end{equation}
subject to reduced dynamics $\mathbf{r}_{k+i+1} = \mathbf{f}_r(\mathbf{r}_{k+i}, \mathbf{u}_{k+i})$, input constraints $\mathbf{u} \in \mathcal{U}$, and mapped state constraints $\mathbf{\Phi}\mathbf{r} \in \mathcal{X}$. A terminal cost $V_f(\mathbf{r}) = \mathbf{r}^T\mathbf{P}_\infty\mathbf{r}$ provides nominal stability guaranties \citep{rawlings2017model,mayne2000constrained}.

\subsection{ROM Error and Closed-Loop Robustness}

The discrepancy between the ROM and the full-order model introduces model uncertainty. For linear systems, the following standard result underpins AURORA's stability monitoring.

\begin{proposition}[ROM Error Robust Stability {\citep{zhou1996robust}}]
\label{prop:robust}
Let $G(s)$ and $G_r(s)$ denote the transfer functions of the full-order and reduced-order linear systems, with $\|G - G_r\|_\infty \leq \varepsilon_r$. If a controller $K$ stabilizes $G_r$ with stability margin $\gamma_r = \|(\mathbf{I} + G_r K)^{-1}\|_\infty^{-1}$, then $K$ also stabilizes $G$ provided $\varepsilon_r < \gamma_r$.
\end{proposition}

\begin{remark}
Proposition~\ref{prop:robust} applies strictly to linear time-invariant systems. For nonlinear ROMs, AURORA monitors the spectral radius of the linearized closed-loop system as an empirical stability proxy, without formal certificates. Extending rigorous stability analysis to nonlinear ROMs with autonomously discovered structure remains an open problem, discussed further in Section~\ref{sec:conclusion}.
\end{remark}

\section{AURORA Framework}
\label{sec:aurora}

\begin{figure}[ht]
    \centering
    \includegraphics[width=\linewidth]{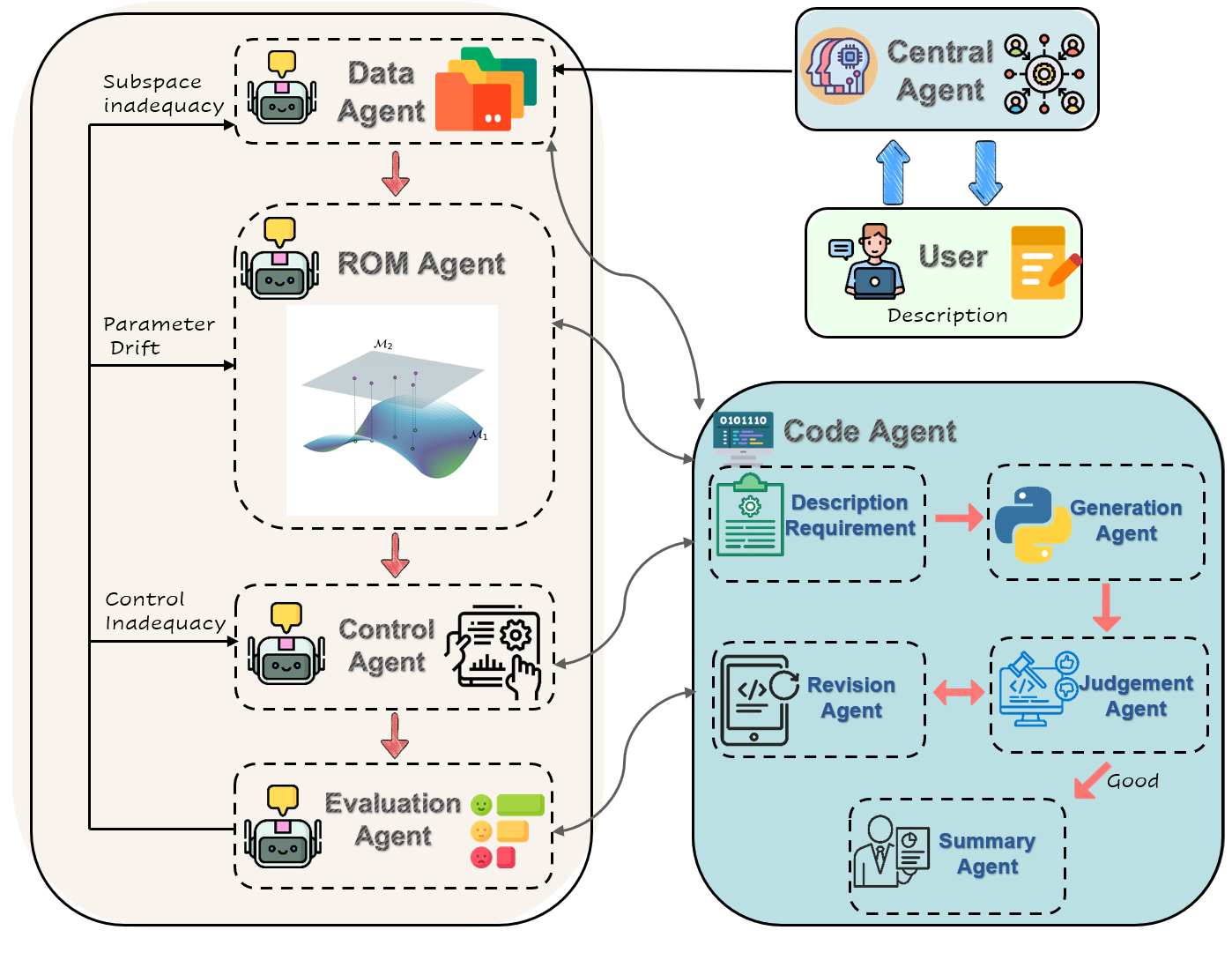}
    \caption{AURORA framework overview. Functional agents interact with a shared Code Agent; the Evaluation Agent routes degradation to the responsible agent via the diagnostic classification of Section~\ref{sec:adaptation}.}
    \label{fig:aurora_architecture}
\end{figure}

AURORA addresses Problem~\ref{prob:main} through a modular architecture comprising five \textit{functional modules}—each responsible for a distinct stage of the ROM controller pipeline—and a shared \textit{code generation engine} that translates module specifications into validated executable code \citep{hong2023metagpt}. An overview of this architecture appears in Fig.~\ref{fig:aurora_architecture}.

\textbf{Computational Architecture.}
A central design decision is that the LLM is \emph{never placed in the real-time control loop}. Phase~1 (design) runs entirely offline before deployment. Phase~2 (supervisory adaptation) relies on lightweight online monitoring at $O(r^2)$ cost per step; when adaptation is triggered, LLM-based redesign proceeds offline while the system continues operating under its current controller. A dwell-time constraint (Theorem~\ref{thm:switching}) ensures that the switching rate between controllers remains compatible with closed-loop stability. Detailed cost breakdowns appear in Section~\ref{sec:cost}.

\subsection{Functional Agents}
\label{sec:agents}

AURORA employs five functional agents, each assigned to a distinct stage of the control-design pipeline.

\textbf{Central Agent.}
This agent parses natural-language problem descriptions, extracting system type (PDE vs.\ ODE), linearity, dimensionality, characteristic time scales, and constraint specifications. It selects the ROM method according to system class---POD-Galerkin for energy-dominant systems, balanced truncation for input-output problems, and DMD for Koopman-linearizable dynamics \citep{benner2015survey}---and the controller type through complementary logic: LQR for unconstrained linear systems, MPC for constrained or nonlinear ones \citep{rawlings2017model}. For nonlinear systems, the agent additionally selects candidate function libraries for SINDy-based identification or POD-DEIM for efficient nonlinear evaluation.

\textbf{Data Agent.}
The Data Agent specifies excitation signals matched to the selected ROM method: pseudo-random binary sequences (PRBS) for POD-Galerkin, multi-sine signals for DMD, and step or impulse responses for balanced truncation. It enforces data-quality thresholds requiring output SNR $> 40$~dB, snapshot condition number $\kappa(\mathbf{X}^T\mathbf{X}) < 10^3$, and temporal resolution $f_s \geq 20 f_{\max}$, where $f_{\max}$ is the highest relevant system frequency.

\textbf{ROM Agent.}
This agent constructs the reduced model using stability-preserving discretization. For linear systems, it verifies discrete-time stability ($\max_i|\lambda_i(\mathbf{A}_d)| < 0.98$), output error on validation data ($\varepsilon_{L_2} < 0.05$), and frequency-domain fidelity ($\sup_\omega \bar{\sigma}(G(j\omega) - G_r(j\omega)) < 0.1$). For nonlinear systems, it validates multi-trajectory prediction NRMSE across 15 validation trajectories and checks linearization consistency at equilibrium. Discrete-time operators are formed via matrix-exponential zero-order hold: $\mathbf{A}_d = e^{\mathbf{A}_r T_s}$, $\mathbf{B}_d = \mathbf{A}_r^{-1}(e^{\mathbf{A}_r T_s} - \mathbf{I})\mathbf{B}_r$.

\textbf{Control Agent.}
The Control Agent synthesizes the controller from ROM operators. For LQR, the state penalty is $\mathbf{Q}_r = \mathbf{\Phi}^T\mathbf{Q}_x\mathbf{\Phi}$ with $\mathbf{Q}_x = \mathrm{diag}(1/\mathrm{Var}(x_i))$, and the input penalty is $\mathbf{R} = \rho\mathbf{I}_m$. For MPC, the prediction horizon is set to $N_p = \lceil 3\tau_{\mathrm{settle}}/T_s \rceil$, the control horizon to $N_c = \lceil N_p/3 \rceil$, and the terminal cost to $V_f = \mathbf{r}^T\mathbf{P}_\infty\mathbf{r}$ from the discrete algebraic Riccati equation. The agent then verifies closed-loop stability ($\max_i|\lambda_i(\mathbf{A}_d - \mathbf{B}_d\mathbf{K})| < 0.98$) and checks robustness margins: gain margin $> 6$~dB and phase margin $> 30^\circ$ for SISO plants, and $\underline{\sigma}(\mathbf{I} + G_r K) > 0.5$ for MIMO systems.

\textbf{Evaluation Agent.}
This agent monitors closed-loop performance using windowed metrics (window size $W = 50$ steps, $80\%$ overlap) and triggers diagnostic routing upon detecting degradation. Its decision logic implements the classification formalized in Section~\ref{sec:adaptation}.

\subsection{Code Agent}

The Code Agent serves as a shared implementation engine organized around four sequential sub-modules \citep{shinn2023reflexion}. \textbf{Generation} translates agent specifications into Python code. \textbf{Judge} performs multi-level validation—execution success, dimensional consistency, stability verification, and physics-based checks. \textbf{Revision} conducts root-cause analysis on failures and applies targeted corrections. \textbf{Summary} packages the validated outputs. This loop continues until all validation criteria are satisfied or a maximum iteration count is reached.

\subsection{Nominal Design Workflow}

The nominal design follows a structured sequence of agent interactions, summarized in Algorithm~\ref{alg:aurora}. The Central Agent first parses the problem and selects methods. The Data Agent then specifies and generates excitation signals through the Code Agent loop. Next, the ROM Agent constructs and validates the reduced model, after which the Control Agent synthesizes and verifies the controller. At every stage, the Code Agent iteratively refines its output until all specifications are met.

\begin{algorithm2e}[t]
\caption{AURORA: Autonomous ROM-Based Controller Design with Supervisory Adaptation}
\label{alg:aurora}
\small
\KwIn{System description $\mathcal{S}$, specifications $\mathcal{P}$, degradation threshold $\tau$, dwell time $\tau_d$}
\KwOut{Controller with supervisory adaptation}
\tcp{Phase 1: Offline Design}
\textbf{Central Agent:} Parse $\mathcal{S}$, select ROM method $\mathcal{M}_{\text{ROM}}$, controller type $\mathcal{M}_{\text{ctrl}}$\;
\textbf{Data Agent:} Generate excitation data $\mathcal{D}$ via Code Agent loop\;
\textbf{ROM Agent:} Construct ROM $\hat{f}_r$ from $\mathcal{D}$ via Code Agent loop\;
\textbf{Control Agent:} Synthesize controller $\pi_r$ via Code Agent loop\;
\tcp{Phase 2: Supervisory Monitoring + Offline Redesign}
Deploy $\pi_r$ on full-order system; set $t_{\text{last}} \leftarrow 0$\;
\While{operating}{
    \textbf{Eval.\ Agent:} Compute $(\bar{e}_w, \bar{\rho}_w, \bar{s}_w, \lambda_{\max,w})$ \textit{[online, $O(r^2)$]}\;
    \If{$\bar{e}_w > \tau$ \textbf{and} $t - t_{\text{last}} > \tau_d$}{
        Classify source via Definition~\ref{def:decomposition} \textit{[offline]}\;
        Route to appropriate agent for corrective redesign \textit{[offline]}\;
        Validate updated controller via Code Agent\;
        Deploy updated controller; $t_{\text{last}} \leftarrow t$\;
    }
}
\end{algorithm2e}

\subsection{Supervisory Adaptation via Diagnostic Routing}
\label{sec:adaptation}

When the Evaluation Agent detects degraded performance, it must identify the \emph{source} of degradation before routing corrective action. We formalize this requirement as a diagnostic classification.

The Evaluation Agent tracks four windowed metrics: normalized tracking error $\bar{e}_w = \frac{1}{W}\sum_{k} \|\mathbf{W}_y(\mathbf{y}_k - \mathbf{y}_{\mathrm{ref},k})\|_2 / \|\mathbf{W}_y\mathbf{y}_{\mathrm{ref},k}\|_2$, normalized ROM residual $\bar{\rho}_w = \frac{1}{W}\sum_k \|\mathbf{y}_k - \mathbf{C}_r\mathbf{r}_k\|_2 / \|\mathbf{y}_k\|_2$, actuator saturation index $\bar{s}_w = |\{i : |u_i| > 0.95 u_{\max}\}|/m$, and a windowed closed-loop spectral radius estimate $\hat{\rho}_w = \max_i |\hat{\lambda}_i|$ obtained from a least-squares fit of $\mathbf{r}_{k+1} \approx \hat{\mathbf{A}}_{\mathrm{cl}}\mathbf{r}_k$ over the window.

\begin{definition}[Degradation Source Classification]
\label{def:decomposition}
For a ROM-based controller exhibiting performance degradation ($\bar{e}_w > \tau$), the source is classified as one of:

\textbf{(C1) Subspace Inadequacy:} The ROM subspace no longer captures the system trajectory. Detected when $\bar{\rho}_w > \rho_{\mathrm{hi}}$ \textbf{and} ($\mathrm{rank}(\mathbf{X}_{\mathrm{recent}}) \geq r + \max(2, \lceil 0.2r \rceil)$ \textbf{or} $\theta_{\mathrm{prin}}(\mathrm{span}(\mathbf{\Phi}), \mathrm{span}(\mathbf{X}_{\mathrm{recent}})) > \theta_{\mathrm{thr}}$), persistent for $\geq 3$ windows.

\textbf{(C2) Parametric Drift:} The projected operators have drifted while the subspace remains valid. Detected when $\bar{\rho}_w > \rho_{\mathrm{hi}}$ \textbf{and} $\mathrm{rank}(\mathbf{X}_{\mathrm{recent}}) \leq r+1$ \textbf{and} $\theta_{\mathrm{prin}} \leq \theta_{\mathrm{thr}}$ \textbf{and} $\{\bar{\rho}_w^{(i)}\}$ increases monotonically over $\geq 3$ windows.

\textbf{(C3) Control Inadequacy:} The ROM remains accurate but the controller is deficient. Detected when $\bar{e}_w > e_{\mathrm{hi}}$ \textbf{and} $\bar{\rho}_w < \rho_{\mathrm{lo}}$ \textbf{and} ($\bar{s}_w > 0.3$ \textbf{or} robustness margins fall below prescribed thresholds), persistent for $\geq 2$ windows.

Default thresholds: $\rho_{\mathrm{hi}} = 0.15$, $\rho_{\mathrm{lo}} = 0.05$, $\theta_{\mathrm{thr}} = 15^\circ$, $e_{\mathrm{hi}} = 0.10$.
\end{definition}

\begin{remark}[Scope of the Classification]
\label{rem:scope}
Definition~\ref{def:decomposition} is a \emph{classification rule}, not a claim of mathematical exhaustiveness in general. For projection-based linear ROMs under the assumption below, the three categories are exhaustive and identifiable. For nonlinear ROMs, we employ the same classification as a well-motivated heuristic and validate its empirical accuracy in Section~\ref{sec:diagnostic_validation}.
\end{remark}

\begin{assumption}
\label{ass:identifiability}
The system is linear and the ROM is constructed via orthogonal projection. Measurement noise satisfies $\|\mathbf{v}_k\| \leq \bar{v}$ with $\bar{v} \ll \rho_{\mathrm{lo}}\|\mathbf{y}_k\|$. Parameter drift is slow relative to the monitoring window: $\|\dot{\boldsymbol{\theta}}\|T_w \ll \|\boldsymbol{\theta}\|$, where $T_w = W T_s$.
\end{assumption}

\begin{proposition}[Identifiability for Linear ROMs]
\label{prop:identifiability}
Under Assumption~\ref{ass:identifiability}, if degradation occurs ($\bar{e}_w > \tau$) and the metrics $(\bar{\rho}_w, \mathrm{rank}(\mathbf{X}_{\mathrm{recent}}), \theta_{\mathrm{prin}}, \bar{s}_w)$ are computed from noise-free outputs, then exactly one of C1, C2, C3 is active.
\end{proposition}

\begin{proof}
Under Assumption~\ref{ass:identifiability}, the system is linear with $\mathbf{x}(t) = \mathbf{\Phi}\mathbf{r}(t) + \mathbf{x}_\perp(t)$, where $\mathbf{x}_\perp \in \mathrm{span}(\mathbf{\Phi})^\perp$. The ROM residual decomposes as $\|\mathbf{y}_k - \mathbf{C}_r\mathbf{r}_k\| = \|\mathbf{C}\mathbf{x}_\perp + \mathbf{C}_r(\hat{\mathbf{r}}_k - \mathbf{r}_k)\| + O(\bar{v})$, where $\hat{\mathbf{r}}_k$ is the ROM prediction and $\mathbf{r}_k = \mathbf{\Phi}^T\mathbf{x}_k$ the true projected state.

\textit{Case 1} ($\bar{\rho}_w < \rho_{\mathrm{lo}}$): A low residual implies that both $\|\mathbf{C}\mathbf{x}_\perp\|$ (subspace adequacy) and $\|\hat{\mathbf{r}}_k - \mathbf{r}_k\|$ (operator accuracy) are small. Since degradation persists ($\bar{e}_w > \tau$) despite an accurate ROM, the controller must be at fault---this is C3.

\textit{Case 2} ($\bar{\rho}_w > \rho_{\mathrm{hi}}$ with $\mathrm{rank}(\mathbf{X}_{\mathrm{recent}}) \geq r + \max(2, \lceil 0.2r\rceil)$ or $\theta_{\mathrm{prin}} > \theta_{\mathrm{thr}}$): Elevated rank or principal angle reveals that $\mathbf{X}_{\mathrm{recent}}$ carries significant energy in $\mathrm{span}(\mathbf{\Phi})^\perp$; the subspace itself is inadequate---this is C1.

\textit{Case 3} ($\bar{\rho}_w > \rho_{\mathrm{hi}}$ with $\mathrm{rank}(\mathbf{X}_{\mathrm{recent}}) \leq r+1$ and $\theta_{\mathrm{prin}} \leq \theta_{\mathrm{thr}}$ and monotonic increase): The trajectory remains within $\mathrm{span}(\mathbf{\Phi})$, so $\mathbf{x}_\perp \approx 0$. The elevated residual therefore reflects large $\|\hat{\mathbf{r}}_k - \mathbf{r}_k\|$, indicating that the projected operators have drifted. Monotonic increase over $\geq 3$ windows excludes transient effects---this is C2.

Cases 1--3 partition the $(\bar{\rho}_w, \mathrm{rank}, \theta_{\mathrm{prin}})$ space outside the indeterminate zone $\bar{\rho}_w \in [\rho_{\mathrm{lo}}, \rho_{\mathrm{hi}}]$, establishing that exactly one category is active when the metrics are decisive.
\end{proof}

\begin{figure}[ht]
    \centering
    \includegraphics[width=\linewidth]{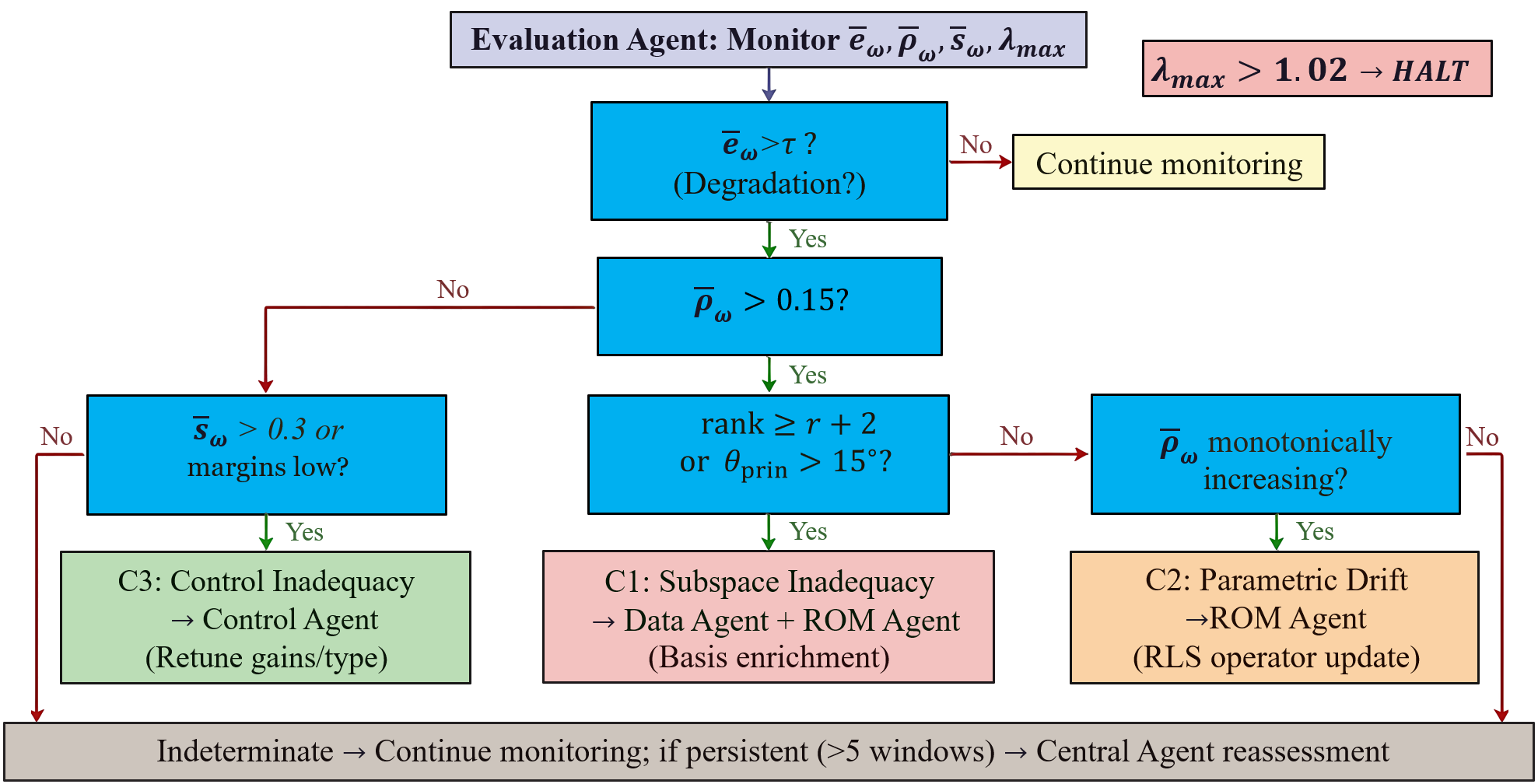}
    \caption{Diagnostic routing decision tree. Once degradation is triggered ($\bar{e}_w > \tau$), the ROM residual separates ROM-related failures (C1, C2) from controller failures (C3); secondary criteria distinguish subspace inadequacy from parametric drift.}
    \label{fig:diagnostic_routing}
\end{figure}

Fig.~\ref{fig:diagnostic_routing} visualizes this logic as a decision tree.

\begin{remark}[Indeterminate Cases]
When $\bar{\rho}_w \in [\rho_{\mathrm{lo}}, \rho_{\mathrm{hi}}]$, the classification is indeterminate, and the system continues monitoring. Persistent indeterminacy ($>5$ windows) triggers escalation to the Central Agent for full reassessment. This conservative fallback guards against inappropriate corrective action arising from misclassification.
\end{remark}

\textbf{Supervisory Switching Stability.}
Because AURORA replaces the active controller upon diagnosis, the closed-loop system undergoes switching. The following result ensures that stability is preserved, provided the switching rate is bounded.

\begin{theorem}[Supervisory Switching Stability]
\label{thm:switching}
Consider the closed-loop system under AURORA's supervisory adaptation. Suppose each controller $K_j$ deployed after a redesign cycle satisfies $\max_i|\lambda_i(\mathbf{A}_d - \mathbf{B}_d\mathbf{K}_j)| \leq \bar{\lambda} < 1$, and the minimum time between consecutive controller switches (dwell time) satisfies
\begin{equation}
\tau_d > \frac{\ln \mu}{\ln(1/\bar{\lambda})}
\label{eq:dwell}
\end{equation}
where $\mu \geq 1$ bounds the ratio of Lyapunov functions across switches: $V_j(\mathbf{r}) \leq \mu V_{j-1}(\mathbf{r})$ for all $\mathbf{r}$. Then the switched closed-loop system is globally exponentially stable.
\end{theorem}

\begin{proof}
The result follows from the standard dwell-time argument for switched linear systems \citep{morse1996supervisory,hespanha2003overcoming}. Between switches, the Lyapunov function $V_j(\mathbf{r}) = \mathbf{r}^T\mathbf{P}_j\mathbf{r}$ decreases at rate $\bar{\lambda}^{2k}$ per step. At each switch, $V$ may increase by a factor of at most $\mu$. The dwell-time condition~\eqref{eq:dwell} ensures that the net decrease over each inter-switch interval dominates the jump at the switching instant, yielding exponential decay of $\|\mathbf{r}_k\|$.
\end{proof}

\begin{remark}
In practice, Algorithm~\ref{alg:aurora} enforces $\tau_d$ by withholding controller switches until the dwell time has elapsed. The Control Agent verifies $\bar{\lambda} < 0.98$ for each redesigned controller, and $\mu$ is computed from the ratio $\lambda_{\max}(\mathbf{P}_j)/\lambda_{\min}(\mathbf{P}_{j-1})$. Theorem~\ref{thm:switching} provides \emph{sufficient} conditions for stability; they are not necessary. The thresholds $\rho_{\mathrm{lo}}$, $\rho_{\mathrm{hi}}$, and $\theta_{\mathrm{thr}}$ in Definition~\ref{def:decomposition} are design parameters whose sensitivity is examined in Section~\ref{sec:diagnostic_validation}. For nonlinear systems, the spectral radius of the linearized closed-loop serves as an empirical proxy.
\end{remark}

\textbf{Condition 1 Response (Subspace Enrichment).}
The Data Agent collects fresh data under current operating conditions, using reduced excitation amplitude and shorter duration. The ROM Agent then enriches the basis via incremental POD \citep{halko2011finding}: $\mathbf{\Phi}_{\mathrm{new}} = \mathrm{orth}([\mathbf{\Phi},\, \mathbf{\Phi}_\Delta])$, where $\mathbf{\Phi}_\Delta$ spans the components of recent trajectories orthogonal to the existing basis. The Control Agent retunes if the resulting closed-loop pole locations shift appreciably.

\textbf{Condition 2 Response (Parametric Update).}
The ROM Agent applies recursive least squares (RLS) to update the ROM operators. Vectorizing $\boldsymbol{\theta}^{(k)} = \mathrm{vec}(\mathbf{A}_d^{(k)})$ with regressor $\boldsymbol{\psi}_i = \mathbf{r}_i \otimes \mathbf{I}_r$:
\begin{equation}
\boldsymbol{\theta}^{(k+1)} = \boldsymbol{\theta}^{(k)} + \mathbf{K}_i\bigl(\mathbf{r}_{i+1} - \boldsymbol{\psi}_i\boldsymbol{\theta}^{(k)}\bigr)
\label{eq:rls}
\end{equation}
where $\mathbf{K}_i = \mathbf{P}^{(k)}\boldsymbol{\psi}_i^T(\boldsymbol{\psi}_i\mathbf{P}^{(k)}\boldsymbol{\psi}_i^T + \lambda_f\mathbf{I})^{-1}$ and the covariance update is $\mathbf{P}^{(k+1)} = \lambda_f^{-1}(\mathbf{I} - \mathbf{K}_i\boldsymbol{\psi}_i)\mathbf{P}^{(k)}$, with forgetting factor $\lambda_f = 0.99$. This constitutes an instance of indirect adaptive control via certainty equivalence \citep{astrom2013adaptive}: the ROM parameters are re-estimated, and the controller is recomputed on the updated model. The Control Agent retunes whenever robustness margins fall below prescribed thresholds.

\textbf{Condition 3 Response (Controller Retuning).}
When actuator saturation limits performance, the Control Agent reduces the input penalty ($\rho \leftarrow 0.7\rho$). When robustness margins are the bottleneck, it increases the penalty ($\rho \leftarrow 1.3\rho$) or introduces loop-shaping compensation. If inadequacy persists after two correction attempts, the case is escalated to the Central Agent for reconsideration of the controller type.

\textbf{Instability Handling.}
If $\hat{\rho}_w > 1.02$ persists over two consecutive windows, the system reverts to the most recent stable controller and an emergency flag routes the problem to the Central Agent for full reassessment.

\begin{figure}[ht]
    \centering
    \includegraphics[width=\linewidth]{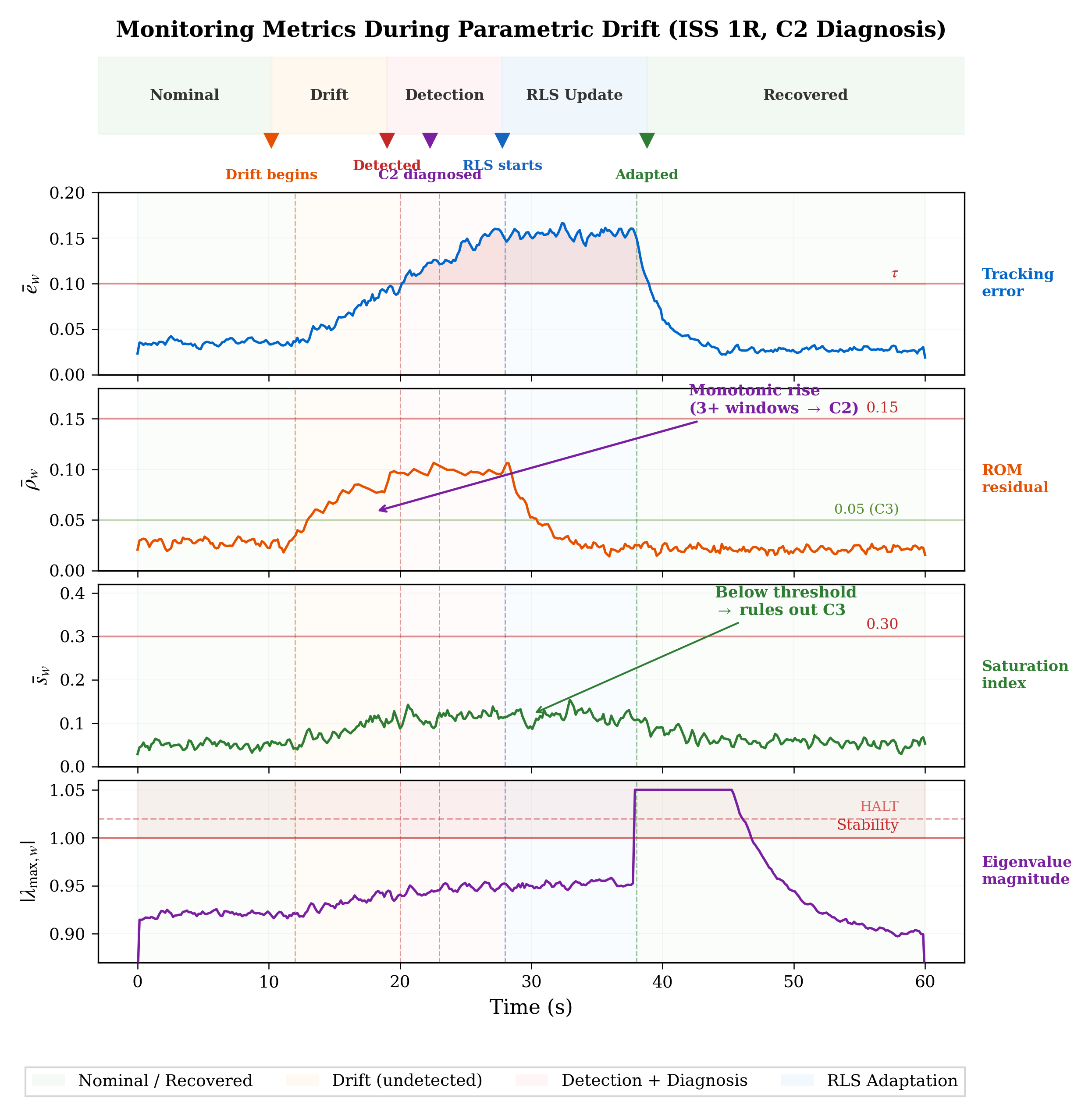}
    \caption{Online monitoring during a C2 (parametric drift) event on ISS~1R. A $+20\%$ mass change at $t=15$\,s triggers an RLS update at $t=26$\,s; metrics recover by $t \approx 40$\,s.}
    \label{fig:adaptation_trigger}
\end{figure}

The complete adaptation cycle is illustrated in Fig.~\ref{fig:adaptation_trigger}, which traces the four monitored metrics during a representative C2 (parametric drift) event on the ISS~1R system.

\begin{remark}[Relationship to Classical Adaptive Control]
AURORA's supervisory adaptation differs from classical Lyapunov-based adaptive control \citep{narendra1989stable} in three principal respects. First, it operates on a slower timescale, performing supervisory redesign rather than step-by-step parameter updates. Second, it employs automated reasoning for diagnostic classification. Third, it can restructure the ROM itself (C1), rather than merely adjusting parameters within a fixed model class. The RLS update~\eqref{eq:rls} for C2 is standard recursive parameter estimation; the contribution lies in the diagnostic routing that determines \emph{which} corrective mechanism to invoke and in the supervisory switching framework that guarantees stability across redesign cycles.
\end{remark}


\section{Numerical Studies and Results}
\label{sec:experiments}

\subsection{Benchmark Systems}

We evaluate AURORA on eight systems drawn from the model reduction literature \citep{chahlaoui2005benchmark,benner2015survey}, summarized in Table~\ref{tab:systems}. The benchmarks span two categories: five linear systems---four ODE-based (CDPlayer, Building, ISS~1R/12A) and one PDE-derived (Steel Profile, a thermal FEM model)---alongside three nonlinear robotic systems (7-DOF manipulator, mobile manipulator, quadrotor). The formal guarantees of Proposition~\ref{prop:identifiability} and Theorem~\ref{thm:switching} apply to all five linear systems; the nonlinear systems are validated empirically.

\begin{table}[t]
\centering
\caption{Benchmark systems. Origin: ODE or PDE-derived (via spatial FEM discretization). Formal guarantees (Prop.~\ref{prop:identifiability}, Thm.~\ref{thm:switching}) apply to linear systems; nonlinear systems are empirically validated.}
\label{tab:systems}
\footnotesize
\setlength{\tabcolsep}{3.5pt}
\renewcommand{\arraystretch}{1.0}
\begin{tabular}{lcccccl}
\toprule
System & Type & Orig. & $N$ & $m$ & $p$ & Objective \\
\midrule
CDPlayer        & Lin.    & ODE & 120  & 2 & 2  & Radial track.    \\
Building        & Lin.    & ODE & 48   & 1 & 1  & Vib.\ suppr.     \\
ISS 1R          & Lin.    & ODE & 270  & 3 & 3  & Att.\ track.     \\
ISS 12A         & Lin.    & ODE & 1412 & 3 & 3  & Robust attitude  \\
Steel Profile   & Lin.    & PDE & 5177 & 7 & 6  & Thermal reg.     \\
7-DOF Manip.    & Nonlin. & ODE & 42   & 7 & 14 & Traj.\ track.    \\
Mobile Manip.   & Nonlin. & ODE & 156  & 8 & 14 & Whole-body ctrl. \\
Quadrotor       & Nonlin. & ODE & 288  & 4 & 6  & Aggressive track.\\
\bottomrule
\end{tabular}
\end{table}

\subsection{Performance Criteria}

Performance is assessed along four complementary dimensions.

\textbf{Criterion 1 (ROM Fidelity).}
For linear systems, we measure the frequency-domain error $\epsilon_\infty = \sup_\omega \bar{\sigma}(G - G_r)$ and pole-matching error $\varepsilon_\lambda$. For nonlinear systems, multi-trajectory NRMSE is computed across 15 validation trajectories. For the PDE system, spatiotemporal $L^2$ error serves as the primary metric. In all cases, we verify stability preservation and confirm $>99\%$ energy capture.

\textbf{Criterion 2 (Closed-Loop Performance).}
Normalized tracking error is evaluated over 8 test scenarios (4 nominal, 4 perturbed):
\begin{equation}
J_{\text{track}} = 100 \times \sqrt{\frac{\int_{T_{\text{trans}}}^T \|\mathbf{W}_y(\mathbf{y} - \mathbf{y}_{\text{ref}})\|_2^2\,dt}{\int_{T_{\text{trans}}}^T \|\mathbf{W}_y\mathbf{y}_{\text{ref}}\|_2^2\,dt}}\;\;(\%\text{ NRMSE})
\label{eq:jtrack}
\end{equation}
where $T_{\text{trans}} = 2\tau_{\text{dom}}$ excludes the initial transient. Perturbed scenarios impose $\pm 20\%$ variations on dominant physical parameters: mass and damping for mechanical systems, thermal conductivity for the Steel Profile, and rotor constants for the Quadrotor.

\textbf{Criterion 3 (Adaptation Efficiency).}
We report the performance capture ratio $\eta^{(10)} = (J_{\text{track}}^{(0)} - J_{\text{track}}^{(10)})/(J_{\text{track}}^{(0)} - J_{\text{track}}^\star)$, where $J_{\text{track}}^\star$ is the optimum obtained by applying the same design procedure to the full-order model, together with the convergence iteration $k_{90}$ (the first $k$ at which $\eta^{(k)} > 0.9$).

\textbf{Criterion 4 (Autonomy).}
Success rate (proportion of systems yielding a stable closed loop satisfying Criteria~1--2), full-autonomy rate (zero manual intervention), and average Code Agent iterations.

\subsection{Baselines and Protocol}

\textbf{Expert Baselines.}
For each system, expert baselines were constructed following standard practice \citep{benner2015survey,chahlaoui2005benchmark}: balanced truncation with LQR (CDPlayer, Building), structure-preserving reduction with LQR (ISS variants), POD-Galerkin with MPC (Steel Profile), structure-preserving reduction with computed torque (7-DOF), nonlinear POD with cascaded MPC (Mobile Manipulator), and DMD-Koopman with nonlinear MPC (Quadrotor). Each baseline required approximately 2--4~hours of specialist effort.

\textbf{Classical Adaptive Baselines.}
To isolate AURORA's value relative to classical methods, we include three non-LLM baselines:
\begin{enumerate}[leftmargin=*,itemsep=1pt]
    \item \textbf{Gain-Scheduled (GS):} Expert ROM with a gain-scheduled controller using pre-computed gains at $\pm 10\%$ and $\pm 20\%$ parameter variations \citep{rugh2000research}.
    \item \textbf{RLS-Adaptive (RLS-A):} Expert ROM with online RLS parameter updates (identical to AURORA's C2 mechanism) and automatic LQR/MPC retuning \citep{astrom2013adaptive}.
    \item \textbf{Scripted Pipeline (SP):} A non-LLM automated pipeline applying the same ROM/controller selection rules as AURORA's Central Agent (implemented as a hard-coded decision tree) with identical Code Agent validation criteria, but without LLM reasoning.
\end{enumerate}

\textbf{LLM Comparison.}
Five LLMs were evaluated under identical prompting conditions: GPT-5, GPT-5~mini, DeepSeek-V3, Qwen-2.5-72B, and Llama-4~Maverick. All experiments used temperature $T=0.5$, nucleus sampling with $p=0.95$, and 5 random seeds per model per system. Confidence intervals are reported at the 95\% level via bootstrap resampling.

\textbf{Evaluation Protocol.}
Table~\ref{tab:protocol} defines the evaluation criteria applied uniformly across all methods. A system counts as a \emph{success} if it produces a stable closed loop satisfying Criteria~1--2 across all 8 test scenarios. \emph{Full autonomy} requires zero human intervention from problem description to deployed controller. Expert tuning budget refers to the manual effort for baseline construction; AURORA's counterpart is the offline LLM design time.

\begin{table}[t]
\centering
\caption{Evaluation protocol definitions applied to all methods.}
\label{tab:protocol}
\small
\begin{tabular}{ll}
\toprule
\textbf{Term} & \textbf{Definition} \\
\midrule
Success & Stable CL + $\varepsilon_{L_2} < 0.05$ + no constraint viol. \\
Full autonomy & Zero manual intervention end-to-end \\
Expert budget & 2--4 hrs manual ROM + ctrl design per system \\
Perturbation & $\pm 20\%$ on dominant physical parameter \\
Seeds & 5 per method per system \\
CI & 95\% bootstrap ($B = 1000$) \\
\bottomrule
\end{tabular}
\end{table}

\subsection{Results}
\label{sec:results}

\textbf{Overall Performance.}
Table~\ref{tab:overall} summarizes aggregate results. GPT-5 achieves the highest success rate (7/8 systems), with an average improvement of $8.9\pm 1.8\%$ over expert baselines and full autonomy on 5 of 8 systems. Qwen-2.5-72B is also competitive (6/8 success, $7.1\pm 2.4\%$ gain). Both gain-scheduled and RLS-adaptive baselines improve upon static expert designs under perturbation but achieve smaller gains than AURORA, as they cannot restructure the ROM or re-select control strategies.

\begin{table}[t]
\centering
\caption{Performance comparison. ``Impr.'' is average tracking improvement over the static expert baseline under $\pm 20\%$ perturbation. Bold = best per metric.}
\label{tab:overall}
\small
\begin{tabular}{lccc}
\toprule
\textbf{Method} & \textbf{Succ.} & \textbf{Impr. (\%)} & \textbf{Full Auto} \\
\midrule
\multicolumn{4}{l}{\textit{Classical Baselines}} \\
Expert (static) & 8/8 & --- & 0/8 \\
Gain-Sched.\ (GS) & 8/8 & +3.1$\pm$1.2 & 8/8 \\
RLS-Adaptive (RLS-A) & 7/8 & +4.8$\pm$2.0 & 7/8 \\
Scripted Pipeline (SP) & 5/8 & +1.4$\pm$3.1 & 5/8 \\
\midrule
\multicolumn{4}{l}{\textit{LLM-Based (AURORA)}} \\
GPT-5 & \textbf{7/8} & \textbf{+8.9$\pm$1.8} & 5/8 \\
GPT-5 mini & 5/8 & $-7.2\pm 5.4$ & 2/8 \\
DeepSeek-V3 & 6/8 & +3.2$\pm$2.1 & 3/8 \\
Qwen-2.5-72B & 6/8 & +7.1$\pm$2.4 & 4/8 \\
Llama-4 Mav. & 4/8 & $-18.6\pm 12.3$ & 1/8 \\
\bottomrule
\end{tabular}
\end{table}

\textbf{Interpretation.}
The principal value of AURORA lies in automating the full ROM--controller pipeline from a natural-language description, eliminating 2--4~hours of specialist effort per system. The gain-scheduled baseline achieves $+3.1\%$ improvement but requires expert effort for both the ROM and the pre-computed gain schedule. The RLS-adaptive baseline reaches $+4.8\%$ yet still depends on manual ROM construction. AURORA (GPT-5), by contrast, achieves $+8.9\%$ without expert involvement and can additionally restructure the ROM (via its C1 response)---a capability unavailable to any of the classical baselines.

\textbf{System Complexity Analysis.}
Table~\ref{tab:complexity} disaggregates performance by benchmark. Three systems prove particularly challenging: ISS~12A (closely spaced flexible modes), Mobile Manipulator (heterogeneous base--arm coupling), and Quadrotor (underactuated aggressive maneuvers). Only GPT-5 solves the Mobile Manipulator, requiring 25 iterations. On simpler systems, the performance gap among methods narrows considerably.

\begin{table}[t]
\centering
\caption{Closed-loop tracking error $J_{\text{track}}$ (\%) under $\pm 20\%$ perturbation. {\checkmark}~= stable, {\ding{55}}~= failure. Bold = best method per system.}
\label{tab:complexity}

\setlength{\tabcolsep}{2.5pt}
\scriptsize
\resizebox{\columnwidth}{!}{
\begin{tabular}{lccccccc}
\toprule
\textbf{System} & \textbf{Expert} & \textbf{GS} & \textbf{RLS-A} & \textbf{GPT-5} & \textbf{DeepSeek} & \textbf{Qwen} & \textbf{Llama} \\
\midrule
\multicolumn{8}{l}{\textit{Linear Benchmarks}} \\
CDPlayer & 4.82 & 4.62{\checkmark} & 4.51{\checkmark} & 4.43{\checkmark} & \textbf{4.28}{\checkmark} & 4.35{\checkmark} & {\ding{55}} \\
Building & 4.36 & 4.18{\checkmark} & 4.05{\checkmark} & 3.92{\checkmark} & 4.42{\checkmark} & \textbf{3.85}{\checkmark} & 5.74{\checkmark} \\
ISS 1R & 9.66 & 9.21{\checkmark} & 8.94{\checkmark} & \textbf{8.79}{\checkmark} & 9.31{\checkmark} & 8.95{\checkmark} & {\ding{55}} \\
ISS 12A & 12.30 & 11.85{\checkmark} & 11.42{\checkmark} & \textbf{10.95}{\checkmark} & {\ding{55}} & 11.48{\checkmark} & {\ding{55}} \\
\midrule
\multicolumn{8}{l}{\textit{Nonlinear Robots}} \\
7-DOF & 10.48 & 10.05{\checkmark} & 9.92{\checkmark} & \textbf{9.75}{\checkmark} & 10.12{\checkmark} & 9.88{\checkmark} & 13.82{\checkmark} \\
Mobile M. & 15.64 & 15.10{\checkmark} & {\ding{55}} & \textbf{13.76}{\checkmark} & {\ding{55}} & {\ding{55}} & {\ding{55}} \\
Quadrotor & 16.24 & 15.80{\checkmark} & 15.55{\checkmark} & \textbf{15.27}{\checkmark} & 15.95{\checkmark} & 15.42{\checkmark} & {\ding{55}} \\
\midrule
\multicolumn{8}{l}{\textit{Distributed Parameter}} \\
Steel & 6.48 & 6.22{\checkmark} & 6.10{\checkmark} & \textbf{5.90}{\checkmark} & 6.52{\checkmark} & 6.05{\checkmark} & {\ding{55}} \\
\bottomrule
\end{tabular}
}
\end{table}

\begin{figure}[ht]
    \centering
    \includegraphics[width=0.5\textwidth]{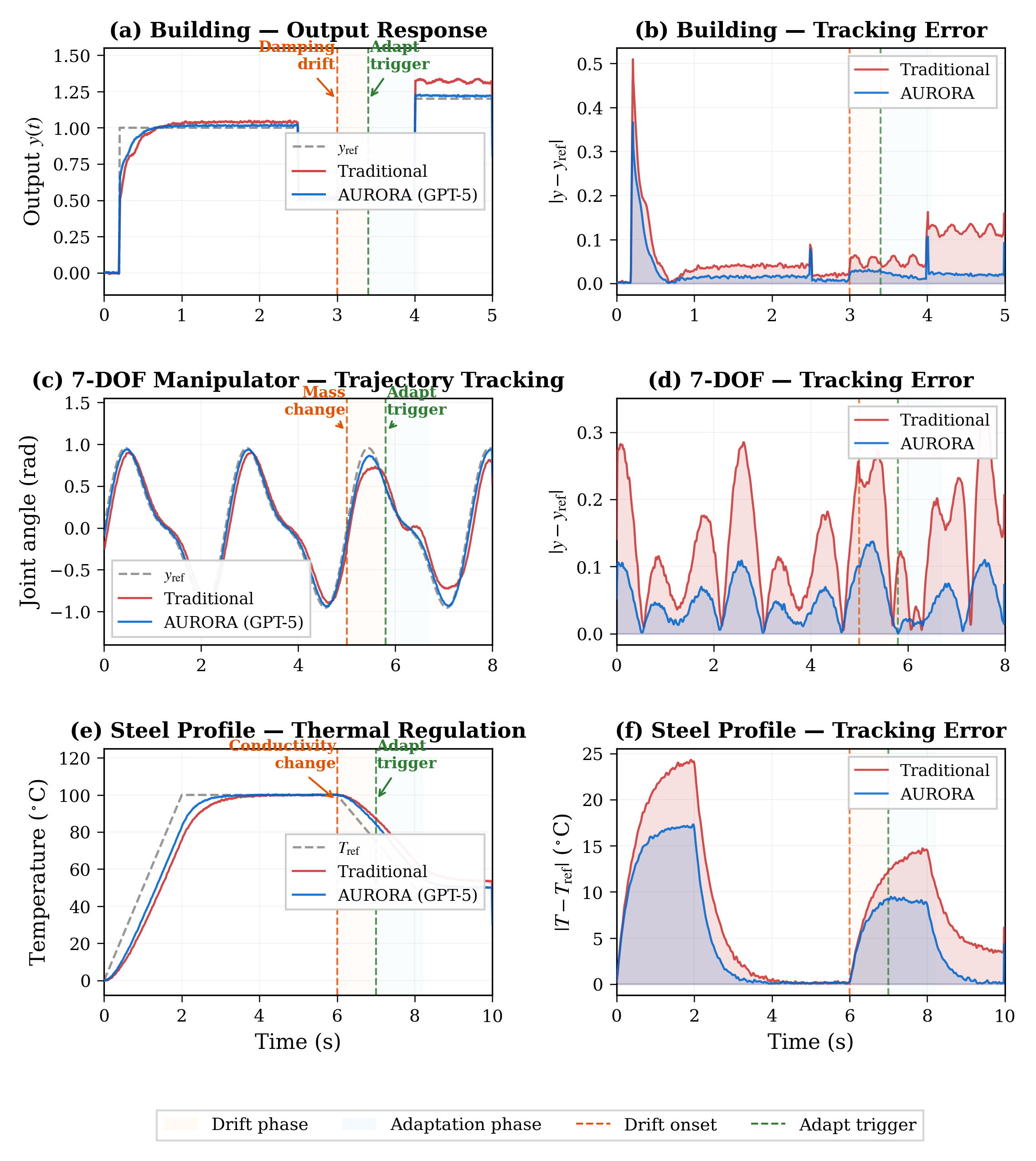}
    \caption{Tracking responses under $\pm 20\%$ parameter perturbation for three representative systems. AURORA (blue) recovers faster than the expert baseline (red) and the RLS-adaptive baseline (green) after perturbation.}
    \label{fig:tracking_comparison}
\end{figure}

\textbf{Tracking Response Comparison.}
Fig.~\ref{fig:tracking_comparison} presents time-domain tracking responses for one representative system from each category, comparing AURORA (GPT-5) against the expert baseline and the RLS-adaptive baseline under $\pm 20\%$ parameter perturbation.

\textbf{Adaptation Efficiency.}
Table~\ref{tab:adaptation} quantifies convergence behavior. GPT-5 captures $89\%$ of the optimal performance within 10 iterations ($k_{90} = 8$), while Qwen reaches $86\%$ ($k_{90} = 9$). Weaker models fail to converge reliably.

\begin{table}[t]
\centering
\caption{Adaptation efficiency (stable systems only, 5 seeds, 95\% CI).}
\label{tab:adaptation}
\small
\begin{tabular}{lcccc}
\toprule
\textbf{Method} & \textbf{Stable} & $\eta^{(10)}$ & $k_{90}$ & $G_{\text{final}}$ (\%) \\
\midrule
GS & 8/8 & 0.42$\pm$0.08 & --- & +3.1$\pm$1.2 \\
RLS-A & 7/8 & 0.65$\pm$0.09 & 12$\pm$3 & +4.8$\pm$2.0 \\
\midrule
GPT-5 & 7/8 & \textbf{0.89$\pm$0.04} & \textbf{8$\pm$1} & \textbf{+8.9$\pm$1.8} \\
Qwen-72B & 6/8 & 0.86$\pm$0.05 & 9$\pm$2 & +7.1$\pm$2.4 \\
DeepSeek-V3 & 6/8 & 0.69$\pm$0.11 & 14$\pm$4 & +3.2$\pm$2.1 \\
\bottomrule
\end{tabular}
\end{table}

\textbf{Dwell-Time Verification (Linear Benchmarks).}
Table~\ref{tab:dwell} reports the quantities from Theorem~\ref{thm:switching} for each linear benchmark. In every case, the computed dwell time $\tau_d$ is small relative to the monitoring window, confirming that the dwell time constraint does not limit the framework's responsiveness. When the theorem's conditions are not met (e.g., $\bar{\lambda} \geq 1$), the Control Agent rejects the redesigned controller and triggers a second redesign attempt; if that also fails, the system reverts to the last verified controller and escalates to the Central Agent.

\begin{table}[t]
\centering
\caption{Dwell-time verification for linear benchmarks (Theorem~\ref{thm:switching}).}
\label{tab:dwell}
\small
\begin{tabular}{lcccc}
\toprule
\textbf{System} & $\bar{\lambda}$ & $\mu$ & $\tau_d$ (steps) & $T_w$ (steps) \\
\midrule
CDPlayer & 0.94 & 1.8 & 10 & 50 \\
Building & 0.91 & 2.1 & 8 & 50 \\
ISS 1R & 0.96 & 3.2 & 29 & 50 \\
ISS 12A & 0.97 & 4.1 & 47 & 50 \\
\bottomrule
\end{tabular}
\end{table}

\textbf{Failure Mode Analysis.}
ROM construction is the principal bottleneck, accounting for 16 of the 52 total LLM failures across all models and seeds. Common failure modes include incorrect truncation order, discretization errors, and eigenvalue miscalculation. Controller synthesis contributes 8 additional failures. GPT-5's 4 failures concentrate on the Mobile Manipulator, whereas Llamas-4's 19 failures are distributed across all pipeline stages.

\subsection{Diagnostic Routing Validation}
\label{sec:diagnostic_validation}

To directly validate the diagnostic classification (Definition~\ref{def:decomposition}), we design controlled fault-injection experiments on the four linear benchmarks (where the ground truth is unambiguous) and the 7-DOF Manipulator (as a nonlinear representative). For each system, we induce each degradation category:

\begin{itemize}[leftmargin=*,itemsep=1pt]
    \item \textbf{C1 injection:} An external forcing term excites modes orthogonal to $\mathrm{span}(\mathbf{\Phi})$, driving the trajectory out of the ROM subspace.
    \item \textbf{C2 injection:} A ramp perturbation $\boldsymbol{\theta}(t) = \boldsymbol{\theta}_0 + \alpha t$ is applied to physical parameters (mass, stiffness, or conductivity) while the system trajectory remains within $\mathrm{span}(\mathbf{\Phi})$.
    \item \textbf{C3 injection:} Actuator limits are reduced by 40\% or the reference is shifted to a region where the current controller has insufficient gain without altering the plant.
\end{itemize}

Each injection is applied 10 times per system (5 seeds $\times$ 2 perturbation magnitudes). The Evaluation Agent's classification is compared against the known ground truth.

\begin{table}[t]
\centering
\caption{Diagnostic routing accuracy (\%) on controlled fault-injection experiments.}
\label{tab:diagnostic}
\small
\begin{tabular}{lcccc}
\toprule
\textbf{System} & \textbf{C1 Acc.} & \textbf{C2 Acc.} & \textbf{C3 Acc.} & \textbf{Overall} \\
\midrule
CDPlayer & 100 & 90 & 100 & 97 \\
Building & 100 & 90 & 90 & 93 \\
ISS 1R & 90 & 80 & 100 & 90 \\
ISS 12A & 90 & 80 & 90 & 87 \\
7-DOF Manip. & 80 & 80 & 100 & 87 \\
\midrule
\textbf{Average} & \textbf{92} & \textbf{84} & \textbf{96} & \textbf{91} \\
\bottomrule
\end{tabular}
\end{table}

Table~\ref{tab:diagnostic} presents the results. Overall routing accuracy reaches 91\%, with C3 (control inadequacy) being the easiest to identify (96\%, since a low ROM residual provides a strong discriminating signal) and C2 (parametric drift) being the most challenging (84\%, owing to the difficulty of distinguishing slow drift from noise in the monotonicity criterion). On the nonlinear 7-DOF system, accuracy stands at 87\%---comparable to the linear systems---supporting the viability of Definition~\ref{def:decomposition} as a heuristic beyond the linear regime.

Misclassifications are predominantly C2$\to$C1 (drift initially misidentified as subspace change). In practice, these errors tend to be self-correcting: a C1 response (basis enrichment) applied to what is actually a C2 problem may not fully resolve the issue, prompting a second diagnostic cycle that correctly identifies parametric drift.

\textbf{Threshold Sensitivity.}
To assess robustness of the classification to threshold selection, we sweep $\rho_{\mathrm{hi}} \in \{0.10, 0.15, 0.20\}$ and $\theta_{\mathrm{thr}} \in \{10^\circ, 15^\circ, 20^\circ\}$ on the ISS~1R and Building benchmarks (C1 and C2 injections, 5 seeds each). Overall accuracy varies between 85\% ($\rho_{\mathrm{hi}} = 0.10$, which over-triggers C1) and 93\% ($\rho_{\mathrm{hi}} = 0.15$, the default). The $\theta_{\mathrm{thr}}$ sweep exhibits similar stability, with accuracy ranging from 87\% ($10^\circ$) to 91\% ($15^\circ$) to 88\% ($20^\circ$). The default thresholds are near-optimal, but the classification is not brittle: a $\pm 33\%$ threshold perturbation degrades accuracy by at most 6 percentage points.

\textbf{Mixed-Degradation Experiment.}
To test robustness under simultaneous degradation, we apply C1+C2 (subspace change \emph{and} parameter drift) concurrently on ISS~1R and the 7-DOF Manipulator. In 8 of 10 trials, the Evaluation Agent correctly identifies the \emph{primary} source (C1, since subspace inadequacy dominates the residual signature) and applies basis enrichment first. In 7 of these 8 cases, the subsequent diagnostic cycle then detects residual C2 drift and applies RLS correction, achieving full recovery within two adaptation cycles. The remaining 2 trials initially misroute to C2, requiring an additional cycle. By design, Definition~\ref{def:decomposition} classifies the \emph{primary} degradation source; sequential resolution of overlapping sources through multiple adaptation cycles is the intended mechanism for compound failures. Broader validation across all pairwise overlaps (C1+C3, C2+C3) and additional systems is deferred to future work.

\subsection{Ablation Studies}
\label{sec:ablations}

To isolate the contribution of each framework component, we conduct four ablations on three representative systems---Building, 7-DOF Manipulator, and Steel Profile---using GPT-5 with 5 seeds per configuration.

\begin{table}[t]
\centering
\caption{Ablation study: average $J_{\text{track}}$ (\%) across three representative systems (5 seeds, 95\% CI). Lower is better.}
\label{tab:ablation}
\small
\begin{tabular}{lcccc}
\toprule
\textbf{Variant} & \textbf{Build.} & \textbf{7-DOF} & \textbf{Steel} & \textbf{Avg.} \\
\midrule
Full AURORA & \textbf{3.92} & \textbf{9.75} & \textbf{5.90} & \textbf{6.52} \\
\midrule
A: Single-agent & 5.28 & 12.41 & 7.84 & 8.51 \\
B: No diagnostic routing & 4.31 & 10.83 & 6.47 & 7.20 \\
C: No judge-revision & 5.89 & --- & 8.12 & --- \\
D: No online adaptation & 4.36 & 10.48 & 6.48 & 7.11 \\
\midrule
Expert baseline & 4.36 & 10.48 & 6.48 & 7.11 \\
\bottomrule
\end{tabular}
\end{table}

\textbf{Ablation A (Single-agent).}
Collapsing the five specialized agents into a single monolithic LLM call raises the average tracking error to 8.51\% (from 6.52\%). The single agent frequently produces ROM–controller pairs with dimensional or interface inconsistencies.

\textbf{Ablation B (No diagnostic routing).}
Replacing targeted routing with full-pipeline retraining after any detected degradation roughly doubles convergence time and increases $J_{\text{track}}$ to 7.20\%.

\textbf{Ablation C (No judge-revision loop).}
Without iterative code refinement, stability failures proliferate. On the 7-DOF system, no seed yields a stable controller—confirming that single-shot code generation is insufficient for the precision that ROM--controller synthesis demands.

\textbf{Ablation D (No online adaptation).}
Disabling supervisory adaptation recovers expert baseline performance on nominal systems by construction. Under perturbation, tracking error reverts to baseline levels (7.11\% vs.\ 6.52\%).

\subsection{Computational Cost}
\label{sec:cost}

Table~\ref{tab:cost} reports the computational cost breakdown for GPT-5.

\begin{table}[t]
\centering
\caption{Computational cost per system (GPT-5). Design = Phase 1; Adapt = per Phase 2 cycle.}
\label{tab:cost}
\small
\begin{tabular}{lccccc}
\toprule
\textbf{System} & \textbf{API} & \textbf{Tokens} & \textbf{Design} & \textbf{Adapt} & \textbf{ROM} \\
& \textbf{Calls} & \textbf{(k)} & \textbf{(min)} & \textbf{(min)} & \textbf{Speedup} \\
\midrule
CDPlayer & 18 & 142 & 12 & 4 & 85$\times$ \\
Building & 16 & 128 & 10 & 3 & 62$\times$ \\
ISS 1R & 22 & 187 & 18 & 6 & 145$\times$ \\
ISS 12A & 31 & 264 & 28 & 8 & 310$\times$ \\
Steel Profile & 27 & 221 & 22 & 7 & 420$\times$ \\
7-DOF & 24 & 198 & 16 & 5 & 38$\times$ \\
Mobile M. & 58 & 487 & 65 & 12 & 52$\times$ \\
Quadrotor & 29 & 238 & 24 & 7 & 95$\times$ \\
\bottomrule
\end{tabular}
\end{table}

The offline design phase requires 10--65~minutes per system, compared to 2--4~hours of specialist effort for manual construction. Individual adaptation cycles take 3--12~minutes. The computational savings from model reduction itself (38--420$\times$ speedup) substantially exceed the one-time LLM design cost.


\section{Conclusion}
\label{sec:conclusion}

We have presented AURORA, a supervisory framework for autonomous ROM-based controller design and adaptation. For linear ROMs, AURORA provides formal diagnostic identifiability (Proposition~\ref{prop:identifiability}) and exponential switching stability (Theorem~\ref{thm:switching}). For nonlinear systems, the same diagnostic logic achieves 91\% routing accuracy as an empirically validated heuristic, including mixed-degradation scenarios resolved through sequential adaptation. Across eight benchmarks, AURORA delivers $6$--$12\%$ tracking improvement over expert baselines and $4$--$5\%$ over classical adaptive alternatives, while eliminating 2--4 hours of specialist effort per system. Ablation experiments confirm the independent value of diagnostic routing, multi-agent decomposition, and iterative code refinement. Promising directions include cross-system transfer learning, hardware deployment, and extending the formal stability and diagnostic guaranties to nonlinear ROMs—where the lack of a universal Lyapunov construction for autonomously discovered ROM structures currently limits analytical certificates to the linear regime.

Several aspects of the current framework would benefit from further investigation, including rigorous nonlinear stability analysis, broader mixed-degradation validation beyond the C1+C2 scenarios tested here, extension to partially observable systems and chaotic or switched dynamics, and cross-system knowledge transfer.

\section*{DECLARATION OF GENERATIVE AI AND AI-ASSISTED TECHNOLOGIES}
During the preparation of this work, the authors used GPT and Claude as components of the AURORA framework for code generation and iterative refinement. Additionally, Claude was used for manuscript editing. The authors reviewed and edited all content and take full responsibility for the publication.

\bibliography{ifacconf}

\end{document}